\documentclass[aps,preprint]{revtex4}%
\usepackage{amsfonts}
\usepackage{amsmath}
\usepackage{amssymb}
\usepackage{graphicx}%
\begin{document}
\preprint{ }
\title{Sigma model effective action for strong localization. }
\author{{\large A. Babichenko}}
\affiliation{\textit{Department of Particle Physics, Weizmann Institute of Science,
\ \ \ \ Rehovot \ 76100, Israel; }}
\affiliation{also at the Department of Applied Mathematics, Holon Institute of Technology,
Holon 58102, Israel.}
\author{{\large V. Babichenko}}
\affiliation{\textit{R.S.C., Kurchatov Institute, Kurchatov sq. 1, Moscow 123182,
\ \ \ \ Russia; \ \ e-mail: vsbabichenko@hotmail.com}}
\pacs{PACS number}

\begin{abstract}
Bose gas in a random external field is considered. The sigma model like
effective action both for weak and strong random fields compared with the
interaction between particles is derived by averaging over the random field
and integration over the quantum fluctuations in the framework of the
Keldysh-Schwinger technique for irreversible processes. Using this effective
action the properties of the system in strongly fluctuating random field are analyzed.

\end{abstract}
\volumeyear{ }
\volumenumber{ }
\issuenumber{number}
\eid{identifier}
\startpage{0}
\endpage{ }
\maketitle

\section{Introduction}

The problem of transition between localized and extended states of many
particle systems attracts attention of condensed matter theorists and
experimentalists for a long time. The theory of non interacting electrons in
the random external field in one and two dimensions predicts localization of
particles for arbitrarily weak disorder \cite{Berez}, \cite{And1}, \cite{Lee}.
In three dimensions the state of non interacting electrons is localized for
sufficiently strong disorder \cite{And2}, \cite{Lif}, and the state is
extended for the disorder less than a critical edge. For negative energy with
absolute value much larger than the mobility edge, the system is strongly
localized and its density of states is described by so called "tail" of
density of states \cite{Lif}, \cite{Zit}, \cite{Halp}, \cite{Lif1}. The common
point of view is that the repulsive interaction between electrons suppresses
localization. So far this problem was mainly considered in the case of week
disorder \cite{Alt}, \cite{Fink}. However the problem of investigation of
transition between localized and extended states caused by interaction between
particles in three dimension is of great importance and interest, but it
cannot be solved in the framework of the weak disorder.

The problem of localization in a random field was investigated for the case of
the Fermi statistics of particles. In this case if the Fermi energy is large
enough the interaction between particles can be neglected, as it was done in
\cite{Berez}, \cite{And1}, \cite{Lee}, \cite{Weg}, \cite{Gor}, \cite{Ef}. The
analysis of the problem when both random field and the interaction between
particles are involved, is the sufficiently more complicated.

In this paper we consider the case of Bose statistics of particles, although,
the technique developed here, in general points, can be applied for the Fermi
case too. For Bose system in random field the interaction between particles
plays fundamental role and cannot be neglected. This is because the nonzero
energy of the Bose system without random potential is defined by the
interaction between particles, and exactly this interaction prevents the fall
of the system to the very bottom of the fluctuating random potential wells.
Our goal is development of a technique for the description of the transition
from the localized state to the delocalized superfluid state. The experimental
and theoretical investigations of this transition in the systems of different
nature are of an interest for a long time \cite{Reppy} - \cite{Billy},
\cite{Fisher}, \cite{Huang}, \cite{Vin}, \cite{Grah}, \cite{Pit}, \cite{Pokr},
\cite{Pokr1}. In the works \cite{Huang} - \cite{Pit} the Bose gas in a small
random external field was considered and corrections to the temperature of the
superfluid transition and to the superfluid density were calculated using
perturbation theory in random field. In the recent works \cite{Pokr},
\cite{Pokr1} the case of the strong random field, or small density of Bose
gas, were considered qualitatively. The technique developed here enables to
consider both strong and weak random fields compared to the strength of
interaction between the Bose particles. The idea to average over the random
field at the first step of calculations if the Keldysh-Schwinger technique
\cite{Keld}, \cite{Schw} was already exploited previously \cite{BK},
\cite{Kam}. We develop this idea and obtain the effective action in the form
of a sigma model, which describes both the cases of the weak and strong random
field. In this work we mainly focus on the application of the obtained
effective action to the case of the strong random field.

\section{Averaged generating functional}

The Keldysh-Schwinger technique \cite{Schw}, \cite{Keld}, along with
supersymmetry and replica technique, is known as an effective in the problems
of localization in a random static potential \cite{BK}, \cite{Kam}. In this
technique absence of denominator in the representation of generating
functional makes it easy to take an average over the Gaussian random static
external potential $V\left(  \overrightarrow{r}\right)  $, provided some
natural assumptions are made, like absence of interactions for asymptotic
in-states which are turned on adiabatically. The generating functional for the
system in random static potential $V\left(  \overrightarrow{r}\right)  $ is%

\begin{equation}
Z\left[  J,V\right]  =\int D\phi D\overline{\phi}\exp\left\{  iS\left[
V\right]  +iS_{J}\right\}  \label{Z}%
\end{equation}
with the action%
\begin{equation}
S\left[  V\right]  =S_{0}+\Delta S_{0}\left[  V\right]  +S_{int} \label{S}%
\end{equation}
and interaction with the sources $S_{J}$. Here $S_{0}$ is the free particle
action
\begin{equation}
S_{0}=%
{\displaystyle\oint}
dtd^{d}r\left\{  \overline{\phi}\left[  i\partial_{t}-\xi\left(
\widehat{\overrightarrow{p}}\right)  \right]  \phi\right\}  , \label{S0}%
\end{equation}
$\Delta S_{0}\left[  V\right]  $ is the interaction of the particles with a
random external field
\begin{equation}
\Delta S_{0}\left[  V\right]  =-%
{\displaystyle\oint}
dtd^{d}r\left\{  \overline{\phi}V\phi\right\}  , \label{S0V}%
\end{equation}
and $S_{int}$ and $S_{J}$ are the interaction between the particles and their
interaction with the external source fields, which are auxiliary and put to
zero at the end of calculations.%
\begin{equation}
S_{int}=-\frac{1}{2}g%
{\displaystyle\oint}
dtd^{d}r\left\{  \overline{\phi}\left(  t,\overrightarrow{r}\right)
\phi\left(  t,\overrightarrow{r}\right)  \overline{\phi}\left(
t,\overrightarrow{r}\right)  \phi\left(  t,\overrightarrow{r}\right)
\right\}  . \label{Sint}%
\end{equation}

\begin{equation}
S_{J}=%
{\displaystyle\oint}
dtd^{d}r\left[  \overline{J}\phi+\overline{\phi}J\right]  \label{SJ}%
\end{equation}
In (\ref{S0})-(\ref{SJ}) the time $t$ integration contour is the standard
Keldysh-Schwinger double time contour with the return, going from $t=-\infty$
to $t=+\infty$ (shifted up) and back to $t=-\infty$ (shifted down) in the
complex $t$ plane. The complex particle fields $\phi$, $\overline{\phi}$ and
their sources $J$, $\overline{J}$\ are functions of the time and space
coordinates $t,\overrightarrow{r}$. At the moment we don't fix the statistics
of the particles, but in what follows we concentrate on the Bose particles
case.\ $g$ is coupling constant, and the function $\xi\left(  \widehat
{\overrightarrow{p}}\right)  $ is the excitation spectrum of the ideal gas,
$\xi\left(  \widehat{\overrightarrow{p}}\right)  =\frac{\overrightarrow{p}%
^{2}}{2m}-\mu$, where the momentum operator is $\widehat{\overrightarrow{p}%
}=-i\overrightarrow{\nabla}$ , and $\mu$ is the chemical potential. We choose
the system of units in which the Plank constant $\hbar$ and the mass of
particles $m$ are equal to unity $\hbar=m=1$.

The probability of the static random field distribution is supposed to be
Gaussian with the correlator of white-noise type%

\[
<V\left(  \overrightarrow{r}\right)  V\left(  \overrightarrow{r}^{\prime
}\right)  >=\left(  1/K_{0}\right)  \delta\left(  \overrightarrow
{r}-\overrightarrow{r}^{\prime}\right)
\]
with some constant $K_{0}$:%

\begin{equation}
P\left[  V\right]  =\frac{\exp\left\{  -\frac{1}{2}K_{0}%
{\displaystyle\int}
d^{d}r\left(  V\left(  \overrightarrow{r}\right)  \right)  ^{2}\right\}
}{\int DV\exp\left\{  -\frac{1}{2}K_{0}%
{\displaystyle\int}
d^{d}rd^{d}r^{\prime}\left(  V\left(  \overrightarrow{r}\right)  \right)
^{2}\right\}  } \label{P}%
\end{equation}
As we said, the absence of denominator in (\ref{Z}) makes it possible to
integrate over the random potential $V$ in the averaged generating functional:%

\begin{equation}
<Z\left[  J,V\right]  >_{V}=\int D\phi D\overline{\phi}DV\exp\left\{
iS\left[  V\right]  +iS_{J}\right\}  P\left[  V\right]  =\int D\phi
D\overline{\phi}\exp\left\{  iS\right\}  \label{Za}%
\end{equation}
where%

\begin{equation}
S=S_{0}+S_{int}+S_{av}+S_{J} \label{S1}%
\end{equation}
The parts of the action $S_{0}$, $S_{int}$, $S_{J}$ are given by equations
(\ref{S0}), (\ref{Sint}), (\ref{SJ}), and new additional interaction term
$S_{av}$ generated by random field averaging, has the form%

\begin{equation}
S_{av}=i\frac{1}{2K_{0}}\int d^{d}r\left[
{\displaystyle\oint}
dt\overline{\phi}\left(  t,\overrightarrow{r}\right)  \phi\left(
t,\overrightarrow{r}\right)  \right]  \left[
{\displaystyle\oint}
dt^{\prime}\overline{\phi}\left(  t^{\prime},\overrightarrow{r}\right)
\phi\left(  t^{\prime},\overrightarrow{r}\right)  \right]  \label{Sav}%
\end{equation}

It is convenient to rewrite the action $S$ in the so called three-angle
representation \cite{Keld}, \cite{LLP}, and introduce the fields $\Psi$ and
$\psi$ instead of the fields $\phi_{1}$ and $\phi_{2}$ , where 1, 2 are the
indexes of the upper and lower brunches of the time contour respectively:%

\begin{align}
\Psi &  =\frac{\phi_{1}+\phi_{2}}{2}\text{; \ \ \ \ \ \ \ \ \ \ }\psi=\phi
_{1}-\phi_{2}\text{\ \ \ }\label{3Ang}\\
J  &  =\frac{J_{1}+J_{2}}{2}\text{; \ \ \ \ \ \ \ \ \ \ \ \ }j=j_{1}%
-j_{2}\text{\ \ \ \ \ \ \ \ \ \ \ \ \ \ \ \ \ \ \ \ }\nonumber
\end{align}
In these variables the parts of the action take the form%

\begin{equation}
S_{0}=%
{\displaystyle\int}
dtd^{d}r\left\{  \overline{\psi}\left(  \widehat{G}_{0R}^{-1}\right)
\Psi+\overline{\Psi}\left(  \widehat{G}_{0A}^{-1}\right)  \psi\right\}
\label{S03a}%
\end{equation}

\begin{equation}
S_{int}=-g%
{\displaystyle\int}
dtd^{d}r\left\{  \left(  \overline{\Psi}\psi+\overline{\psi}\Psi\right)
\left(  \overline{\Psi}\Psi+\frac{1}{4}\overline{\psi}\psi\right)  \right\}
\label{Sint3a}%
\end{equation}

\begin{equation}
S_{av}=\frac{i}{2K_{0}}\int d^{d}r\left\{  \int dt\left[  \overline{\Psi}%
\psi+\overline{\psi}\Psi\right]  \right\}  \left\{  \int dt^{\prime}\left[
\overline{\Psi}\psi+\overline{\psi}\Psi\right]  \right\}  \label{Sav3a}%
\end{equation}

\begin{equation}
S_{J}=\int dtd^{d}r\left\{  \left(  \overline{J}\psi+\overline{\psi}J\right)
+\left(  \overline{j}\Psi+\overline{\Psi}j\right)  \right\}  \label{SJ3a}%
\end{equation}
where the inverse retarded and advanced free field Green functions
$\widehat{G}_{0R}$, $\widehat{G}_{0A}$ in frequency representation are%

\begin{align}
\left(  \widehat{G}_{0R}^{-1}\right)   &  =\omega+\mu+\frac{1}{2m}%
\overrightarrow{\nabla}^{2}+i\gamma\label{G0RA}\\
\left(  \widehat{G}_{0A}^{-1}\right)   &  =\omega+\mu+\frac{1}{2m}%
\overrightarrow{\nabla}^{2}-i\gamma\nonumber
\end{align}
with infinitesimal positive $\gamma$. In time representation $\omega$ should
be replaced by $i\partial_{t}$. In the three-angle representation asymptotic
of the fields $\psi$, $\overline{\psi}$ at $t\rightarrow\infty$ satisfy the
condition $\psi\left(  t\rightarrow\infty,\overrightarrow{r}\right)
=\overline{\psi}\left(  t\rightarrow\infty,\overrightarrow{r}\right)  =0$, and
the fields $\Psi$, $\overline{\Psi}$ are non zero at $t\rightarrow\infty$
\cite{BK}. Note that the classical fields, i.e. the fields which in functional
integral technique correspond not to operators but to c-number values, have
$\psi=\phi_{1}-\phi_{2}$ equal to zero at any time moment. The fields $\psi$,
$\overline{\psi}$ describe quantum fluctuations of the quantum system, and
$\Psi$, $\overline{\Psi}$ can be considered as the classical components of the
fields \cite{BK}, \cite{Kam}. In the case of large occupation of some degrees
of freedom the quantum fluctuations $\psi$, $\overline{\psi}$ corresponding to
these degrees of freedom give small corrections compared to the classical
components $\Psi$, $\overline{\Psi}$, and in principle can be taken into
account in the lower orders of perturbation theory. If the system has Bose
condensate it is described by the fields $\Psi$, $\overline{\Psi}$.

\section{Integration over quantum fluctuations.}

Application of the Keldysh-Schwinger technique for localization problem in
Fermi systems in weak random field \cite{Kam} was based on the construction of
so called Q-Lagrangian \cite{Weg}, like for example in replica or
supersymmetry technique \cite{Weg}-\cite{Ef1}. Correctness of application of
this technique to strong localization problem is doubtful. The main goal of
this paper is to consider the properties of many-particle system in strong
random field. Here we consider Bose gas of small density in strong random
field. We suppose $ng^{3}<<1$, where $n$ is the averaged density of the gas,
$g$ is related to the scattering length \cite{Bel}, and the fluctuations of
random field are assumed to be larger or of the order of $\mu_{r}=ng$.

The novelty of our approach is that we start from integration over the quantum
fields $\psi$, $\overline{\psi}$, getting an effective action for the
classical fields $\Psi$, $\overline{\Psi}$. In order to integrate over $\psi$,
$\overline{\psi}$ we introduce new fields $\eta$, $\overline{\eta}$ by%

\begin{align}
\psi\left(  t,\overrightarrow{r}\right)   &  =\frac{1}{\left(  \overline{\Psi
}\left(  t,\overrightarrow{r}\right)  \Psi\left(  t,\overrightarrow{r}\right)
\right)  }\Psi\left(  t,\overrightarrow{r}\right)  \eta\left(
t,\overrightarrow{r}\right)  \label{Et}\\
\overline{\psi}\left(  t,\overrightarrow{r}\right)   &  =\overline{\eta
}\left(  t,\overrightarrow{r}\right)  \overline{\Psi}\left(  t,\overrightarrow
{r}\right)  \frac{1}{\left(  \overline{\Psi}\left(  t,\overrightarrow
{r}\right)  \Psi\left(  t,\overrightarrow{r}\right)  \right)  }\nonumber
\end{align}
We preserved the form of equations which will be valid for both statistics of
the fields $\psi,\Psi$, although our goal in this paper is, as we said, the
bosonic case. In this context the above equations are written in the way which
avoids a division by a Grassmann variable in the fermionic case. Here and
below the form of equation (\ref{Et}) is also convenient in the case of
additional degrees of freedom for the field $\Psi$, like spin. The action
$S_{av}$ is Gaussian in $\eta$, $\overline{\eta}$. Our strategy will be to
integrate over the fields $\eta$, $\overline{\eta}$ and write the effective
action in terms of the chiral fields. Without the interaction term $S_{int}$
this integration is exact both for Fermi and Bose cases. The effective action
takes the form of a sigma model. In this paper the effective action will be
obtained for the Bose gas with the repulsive interaction between particles,
but the technique used here works in fermionic case as well.

The Jacobian of transformations (\ref{Et}) is%

\begin{equation}
\det\left(  \frac{\delta\psi,\delta\overline{\psi}}{\delta\eta,\delta
\overline{\eta}}\right)  =%
{\displaystyle\prod\limits_{t,\overrightarrow{r}}}
\frac{1}{\left(  \overline{\Psi}\Psi\right)  } \label{Det}%
\end{equation}
In terms of new variables $\eta,\overline{\eta}$ the action $S_{av}$ takes the form%

\begin{equation}
S_{av}=\frac{i}{2K_{0}}\int d^{d}r\int dt\int dt^{\prime}\left(
\overline{\eta}\left(  t,\overrightarrow{r}\right)  +\eta\left(
t,\overrightarrow{r}\right)  \right)  \left(  \overline{\eta}\left(
t^{\prime},\overrightarrow{r}\right)  +\eta\left(  t^{\prime},\overrightarrow
{r}\right)  \right)  \label{Sav1}%
\end{equation}
This is the crucial point of the calculation: there are two factorized time
integrations in the last equation.

For future convenience we rewrite the action, introducing new (chiral) fields
$\overline{e},e,\overline{f},f$ instead of the classical fields $\overline
{\Psi},\Psi$ :
\begin{equation}
e=\frac{1}{\sqrt{\overline{\Psi}\Psi}}\Psi\text{; \ \ \ \ \ \ \ \ \ \ }%
\overline{e}=\overline{\Psi}\frac{1}{\sqrt{\overline{\Psi}\Psi}}%
\text{\ \ \ \ \ \ \ \ \ \ \ \ \ \ } \label{e}%
\end{equation}

\begin{equation}
f=\exp\left(  \frac{1}{2}\lambda\right)  \text{; \ \ \ \ \ \ \ \ \ \ \ \ \ }%
\overline{f}=\exp\left(  -\frac{1}{2}\lambda\right)  \text{\ } \label{e1}%
\end{equation}

\begin{equation}
\lambda=\ln\left(  \frac{\overline{\Psi}\Psi}{n_{c}^{\left(  0\right)  }%
}\right)  \text{\ } \label{lambda}%
\end{equation}
The constant $n_{c}^{\left(  0\right)  }$, independent of $\overrightarrow{r}%
$, is introduced to make the argument of logarithm dimensionless. The chiral
fields satisfy the following constraints%

\begin{equation}
\overline{e}e=1\text{, \ \ }\overline{f}f=1\text{\ }\label{chiral}%
\end{equation}
At this stage we will neglect the highest order terms of the quantum
fluctuations $\psi$ in $S_{int}$\ (\ref{Sint3a}) proportional to $\psi^{3}$.
It is quite well understood \cite{Schw}, \cite{Keld}, \cite{LLP}, \cite{Kam}
that these terms are responsible for relaxation processes induced by
interaction between the particles \cite{Bel}. These interaction responsible
terms can be neglected if the density of the system or the interaction between
particles are small $ng^{3}<<1$. Lets emphasis that this approximation,
nevertheless, give the possibility to consider random field fluctuations of
any strength exactly. They are essential if the kinetics of relaxation to an
equilibrium state is considered. In this paper we suppose that the system is
at an equilibrium state and there are no relaxation processes.

With this assumption one can see that in terms of new fields (\ref{e}%
)-(\ref{lambda}) the free and the interaction parts of the action
(\ref{S03a}),(\ref{Sint3a}) combine to%

\begin{equation}
S_{0}+S_{int}=%
{\displaystyle\int}
dtd^{d}r\left\{  \overline{\eta}L+\overline{L}\eta\right\}  \label{S01}%
\end{equation}
where%

\begin{align}
L\left(  t,\overrightarrow{r}\right)   &  =\overline{f}\left(  i\partial
_{t}f\right)  +\overline{e}\left(  i\partial_{t}e\right)  +\mu+\frac{1}%
{2m}\overline{e}\left(  \overrightarrow{\nabla}^{2}e\right)  +\label{L1}\\
&  +\frac{1}{2m}\overline{f}\overrightarrow{\nabla}^{2}f+\frac{1}{2m}\left(
\overrightarrow{\nabla}\lambda\right)  \left(  \overline{e}\overrightarrow
{\nabla}e\right)  -gn_{c}\left(  f\right)  ^{2}\nonumber
\end{align}

\begin{align}
\overline{L}\left(  t,\overrightarrow{r}\right)   &  =\left(  i\partial
_{t}\overline{f}\right)  f+\left(  -i\partial_{t}\overline{e}\right)
e+\mu+\frac{1}{2m}\left(  \overrightarrow{\nabla}^{2}\overline{e}\right)
e+\label{L2}\\
&  +\frac{1}{2m}\overline{f}\overrightarrow{\nabla}^{2}f+\frac{1}{2m}\left(
\overrightarrow{\nabla}\lambda\right)  \left(  \left(  \overrightarrow{\nabla
}\overline{e}\right)  e\right)  -gn_{c}^{\left(  0\right)  }\left(  f\right)
^{2}\nonumber
\end{align}

\begin{align}
L+\overline{L}  &  =\overline{e}\left(  i\partial_{t}e\right)  +\left(
-i\partial_{t}\overline{e}\right)  e+2\mu-\frac{1}{m}\left(  \overrightarrow
{\nabla}\overline{e}\right)  \left(  \overrightarrow{\nabla}e\right)
+\label{L1L2}\\
&  +\frac{1}{m}\overline{f}\overrightarrow{\nabla}^{2}f-2gn_{c}^{\left(
0\right)  }\left(  f\right)  ^{2}\nonumber
\end{align}
The constant $n_{c}^{\left(  0\right)  }$ with dimension of density is
arbitrary and actually drops out from the equation (\ref{L1L2}) because of the
gradient operator and Eq. (\ref{e1}). We will fix its choice below.

Eq. (\ref{L1L2}) was obtained from Eqs. (\ref{L1}, \ref{L2}) using the
constraint (\ref{chiral}) on the fields $e$, $\overline{e}$. The remaining
part of the action $S_{J}$ can be written as%
\begin{equation}
S_{J}=\int dtd^{d}r\left\{  (n_{c}^{\left(  0\right)  })^{-1/2}\left(
\overline{J}e\eta+\overline{\eta}\overline{e}J\right)  \overline{f}%
+(n_{c}^{\left(  0\right)  })^{1/2}f\left(  \overline{j}e+\overline
{e}j\right)  \right\}  \label{SJe}%
\end{equation}
Equations (\ref{e}) can be written in polar coordinates $\rho,\varphi$ for
complex fields $\Psi$,$\overline{\Psi}$:%

\begin{align}
\Psi &  =\rho\exp\left(  i\varphi\right)  =\rho e\text{; \ }\overline{\Psi
}=\rho\exp\left(  -i\varphi\right)  =\rho\overline{e}\text{; \ }%
\label{PsaiRoFai}\\
\rho &  =\sqrt{\overline{\Psi}\Psi}\text{\ }\nonumber
\end{align}
Note that the functional integration measure changes under the transformation
$\Psi$, $\overline{\Psi}\rightarrow\lambda,\varphi$\ as%

\[%
{\displaystyle\prod\limits_{t,\overrightarrow{r}}}
\frac{1}{\left(  \overline{\Psi}\Psi\right)  }D\Psi D\overline{\Psi}D\eta
D\overline{\eta}=%
{\displaystyle\prod\limits_{t,\overrightarrow{r}}}
D\lambda D\varphi D\eta D\overline{\eta}%
\]
If we split out the zero frequency mode (time independent) part $\eta
_{0},L_{0}$ of \ $\eta,L$%

\begin{equation}
\eta\left(  t,\overrightarrow{r}\right)  =\frac{1}{T}\eta_{0}\left(
\overrightarrow{r}\right)  +\delta\eta\left(  t,\overrightarrow{r}\right)
\text{; \ \ \ \ \ \ \ }\overline{\eta}\left(  t,\overrightarrow{r}\right)
=\frac{1}{T}\overline{\eta}_{0}\left(  \overrightarrow{r}\right)
+\delta\overline{\eta}\left(  t,\overrightarrow{r}\right)  \label{Hai}%
\end{equation}%
\[
L\left(  t,\overrightarrow{r}\right)  =L_{0}\left(  \overrightarrow{r}\right)
+\delta L\left(  t,\overrightarrow{r}\right)  \text{; \ \ \ \ \ \ \ }%
\overline{L}\left(  t,\overrightarrow{r}\right)  =\overline{L}_{0}\left(
\overrightarrow{r}\right)  +\delta\overline{L}\left(  t,\overrightarrow
{r}\right)
\]
with the total time evolution $T\rightarrow\infty$, the action $S_{av}$
(\ref{Sav1}) becomes just%

\begin{equation}
S_{av}=\frac{2i}{K_{0}}\int d^{d}r\left(  \eta_{0}\left(  \overrightarrow
{r}\right)  \right)  ^{2} \label{Sav2}%
\end{equation}
whilst the other terms of the action take the form%

\begin{equation}
S_{0}+S_{int}=\int d^{d}r\left(  \frac{\int dt\left(  L_{0}+\overline{L}%
_{0}\right)  }{T}\right)  \eta_{0}+\int d^{d}rdt\left(  \delta\overline{\eta
}\delta L+\delta\overline{L}\delta\eta\right)  \label{S0Sint}%
\end{equation}

Since the last term in (\ref{S0Sint}) is the only one where $\delta\eta$,
$\delta\overline{\eta}$ appears, functional integration over them gives
additional constraint $\delta L=\delta\overline{L}=0$. It means that only time
independent (zero frequency time Fourier components of the) functions $L $,
$\overline{L}$ contribute to the functional integral. However, in general it
doesn't mean that fields $e$, $\overline{e}$, $f$, $\overline{f}$ are time
independent. $L$, $\overline{L}$\ can be taken at any time moment, e.g. at
$t=\infty$. With this choice we will omit the time argument for the functions
$e$, $\overline{e}$, $f$, $\overline{f}$ below.

The first two terms of $L+\overline{L}$ Eq. (\ref{L1L2}) in the first summand
can be transformed to a renormalization of the chemical potential: In terms of
polar variables (\ref{PsaiRoFai}) they are transformed into%

\[
\frac{1}{T}\int d^{d}rdt\left[  \overline{e}\left(  i\partial_{t}e\right)
+\left(  -i\partial_{t}\overline{e}\right)  e\right]  \eta_{0}=-2\int
d^{d}r\frac{\varphi\left(  \infty,\overrightarrow{r}\right)  -\varphi\left(
-\infty,\overrightarrow{r}\right)  }{T}\eta_{0}\left(  \overrightarrow
{r}\right)
\]
The difference of phases $\Delta\varphi\left(  \overrightarrow{r}\right)  =$
$\varphi\left(  \infty,\overrightarrow{r}\right)  -\varphi\left(
-\infty,\overrightarrow{r}\right)  $ can be written as $\Delta\varphi\left(
\overrightarrow{r}\right)  =\delta\varphi\left(  \overrightarrow{r}\right)
+2\pi n$, where $0\leq\delta\varphi\left(  \overrightarrow{r}\right)  <2\pi$
and $n$ is an integer number, which should not depend on the space coordinate
$\overrightarrow{r}$. The changes of $\Delta\varphi\left(  \overrightarrow
{r}\right)  $ by discontinuous jumps means a divergence of its gradient, and
as a consequence to zero contribution to the functional integral. Thus, the
dependence on the constant integer number $n$ can be included into the
renormalization of the chemical potential $\mu$, and the term $\delta
\varphi/T\symbol{126}2\pi/T$ can be neglected for large $T$.

Integration over $\eta_{0}$ in the generating functional $<Z\left[
J,V\right]  >_{V}$ finally gives%

\begin{equation}
<Z\left[  J,V\right]  >_{V}=\int DeDf\exp\left\{  -S_{eff}+iS_{j}\right\}
\label{Zeff}%
\end{equation}
where%

\begin{equation}
S_{eff}=\frac{K_{0}}{8}\int d^{d}r\left[  L\left(  \overrightarrow{r}\right)
+\overline{L}\left(  \overrightarrow{r}\right)  \right]  ^{2} \label{Seff}%
\end{equation}
and $S_{j}$ is also expressed in terms of the fields $e$, $f$. The effective
action $S_{eff}$ can be written in the form%

\begin{equation}
S_{eff}=\frac{1}{2^{1+d/2}}\left(  \frac{|\mu|}{E_{0}}\right)  ^{2-d/2}\int
d^{d}RL_{eff}^{2}(\overrightarrow{R}) \label{Seff1}%
\end{equation}
where $E_{0}$ is the Larkin energy, which characterizes the scale of the
fluctuating random potential $V$.%

\[
E_{0}=\frac{1}{\left(  K_{0}\right)  ^{2/\left(  4-d\right)  }}%
\]

and $L_{eff}(\overrightarrow{R})$ is dimensionless representation of the
function $\frac{1}{2}\left(  L\left(  \overrightarrow{r}\right)  +\overline
{L}\left(  \overrightarrow{r}\right)  \right)  $:%

\begin{equation}
L_{eff}(\overrightarrow{R})=\pm1-\left(  \overrightarrow{\nabla}_{R}%
\overline{e}\right)  \left(  \overrightarrow{\nabla}_{R}e\right)
+\overline{f}\left(  \overrightarrow{\nabla}_{R}\right)  ^{2}f-\frac{g}{|\mu
|}n_{c}^{\left(  0\right)  }\left(  f\right)  ^{2} \label{Leff1}%
\end{equation}
The sign of $\pm1$ here is the sign of the chemical potential $\mu$, and
$\overrightarrow{R}$ is the dimensionless coordinate variable%

\begin{equation}
\overrightarrow{R}=\overrightarrow{r}\sqrt{2|\mu|} \label{R}%
\end{equation}
and the gradient operator is $\overrightarrow{\nabla}_{R}=\frac{\partial
}{\partial\overrightarrow{R}}$. The dimensionless interaction constant
$\alpha=\frac{g}{\mid\mu\mid}n_{c}^{\left(  0\right)  }$ can be fixed to unity
if we choose $n_{c}^{\left(  0\right)  }$ as%

\begin{equation}
n_{c}^{\left(  0\right)  }=\frac{|\mu|}{g} \label{nc}%
\end{equation}
and the Lagrangian $L_{eff}$ takes the form%

\begin{equation}
L_{eff}(\overrightarrow{R})=\pm1-\left(  \overrightarrow{\nabla}_{R}%
\overline{e}\right)  \left(  \overrightarrow{\nabla}_{R}e\right)
+\overline{f}\left(  \overrightarrow{\nabla}_{R}\right)  ^{2}f-\left(
f\right)  ^{2} \label{Leff2}%
\end{equation}

Note that the effective action Eq. (\ref{Seff1}) is derived without any
assumptions about weakness of disorder and can be used for the description of
the transition from strong localization to delocalization and superfluidity.

\section{Calculation of chemical potential.}

In this section, as an application of the derived effective potential, we
calculate the density of particles in three spacial dimensions ($d=3$). The
correlator $<\overline{\Psi}\left(  \overrightarrow{r},t\right)  \Psi\left(
\overrightarrow{r},t\right)  >$ defining the density of the particles, can be
written in the form of functional integral%

\begin{equation}
<\overline{\Psi}\left(  \overrightarrow{r},t\right)  \Psi\left(
\overrightarrow{r},t\right)  >=\int DeD\overline{e}DfD\overline{f}\left(
f\left(  \overrightarrow{r}\right)  \right)  ^{2}\exp\left\{  -S_{eff}%
\right\}  \label{PsiPsi}%
\end{equation}
We consider the case of the large negative chemical potential $\mu<0$,
$\mid\mu\mid>>E_{0}$. The total number of particles $N$ is defined by this correlator%

\begin{equation}
N=\int d^{3}r<\overline{\Psi}\left(  \overrightarrow{r},t\right)  \Psi\left(
\overrightarrow{r},t\right)  > \label{N}%
\end{equation}
and below we calculate the chemical potential $\mu$ as a function of the
average particles density $n=N/\Omega$, where $\Omega$ is the total space
volume of the system.

It is clear that only $e$, $f$ field configurations with non diverging
$L_{eff}$ will contribute to the functional integral (\ref{PsiPsi}). In
particular, it means that $L_{eff}\left(  \overrightarrow{R}\right)  $ should
tend to zero rapidly enough for $\overrightarrow{R}\rightarrow\infty$. Thus,
for large $\overrightarrow{R}$, we have the equation%

\[
-\left(  \overrightarrow{\nabla}_{R}\overline{e}\right)  \left(
\overrightarrow{\nabla}_{R}e\right)  +\overline{f}\left(  \overrightarrow
{\nabla}_{R}\right)  ^{2}f-1-\left(  f\right)  ^{2}=0
\]
Solution of this equation defines the asymptotic $\overrightarrow
{R}\rightarrow\infty$ of an instanton. In the whole region of distances $R$
the instanton is defined by extremum of the action $S_{eff}$. Lets find
solutions with $e=const$. Then the equation takes the form%

\[
\overline{f}\left(  \overrightarrow{\nabla}_{R}\right)  ^{2}f-1-\left(
f\right)  ^{2}=0
\]
Using the constraint $\overline{f}f=1$, we obtain%

\begin{equation}
\left(  \overrightarrow{\nabla}_{R}\right)  ^{2}f-f-\left(  f\right)  ^{3}=0
\label{EqInst2}%
\end{equation}
We will look for solution in the class of spherically symmetric functions of
the form $f=\exp(\frac{1}{2}\lambda)$. Then Laplacian is%

\begin{equation}
\left(  \overrightarrow{\nabla}_{R}\right)  ^{2}=\frac{d^{2}}{dR^{2}}+\frac
{2}{R}\frac{d}{dR} \label{Laplas}%
\end{equation}
Thus, for large $R\rightarrow\infty$ Eq. (\ref{EqInst2}) can be written as%

\begin{equation}
\frac{d^{2}}{dR^{2}}f-f-\left(  f\right)  ^{3}=0 \label{EqInst3}%
\end{equation}
Moreover, we will look for solutions with the asymptotic%

\begin{equation}
f\left(  R\right)  =\exp\left(  -R\right)  \text{ \ \ \ as \ \ \ }%
R\rightarrow\infty\label{e0r}%
\end{equation}
which means that the third term in Eq. (\ref{EqInst3}) can be neglected at
$R\rightarrow\infty$. The center of the instanton is an arbitrary parameter
$\overrightarrow{r}_{0}$, or its dimensionless analog $\overrightarrow{R}%
_{0}=$ $\sqrt{2|\mu|}\overrightarrow{r}_{0}$.%

\begin{equation}
f\left(  \overrightarrow{R}\right)  =\exp\left(  -|\overrightarrow
{R}-\overrightarrow{R}_{0}|\right)  \text{ \ \ \ \ \ \ \ as \ \ }%
\overrightarrow{R}\rightarrow\infty\label{e0R}%
\end{equation}
and the integration over $\overrightarrow{R}_{0}$ is the integration over the
zero mode. All instantons which differ only by parameter $\overrightarrow
{R}_{0}$ give the same contribution to the functional integral, due to the
translational invariance of the effective action.

So we get in quasi classical limit for the correlator%

\begin{align}
&  <\overline{\Psi}\left(  \overrightarrow{r},t\right)  \Psi\left(
\overrightarrow{r},t\right)  >\label{ZM}\\
&  =n_{c}^{\left(  0\right)  }\int d^{3}R_{0}\left(  f^{inst}\left(
\overrightarrow{R}\right)  \right)  ^{2}\exp\left\{  -\frac{1}{2^{5/2}}\left(
\frac{|\mu|}{E_{0}}\right)  ^{1/2}\int d^{3}R\left(  L^{inst}\right)
^{2}\right\} \nonumber
\end{align}
where%
\[
L^{inst}=\overline{f}^{inst}\left(  \overrightarrow{\nabla}_{R}\right)
^{2}f^{inst}-1-\left(  f^{inst}\right)  ^{2}%
\]

One can look for the instanton configuration in the form%

\begin{equation}
f^{inst}\left(  \overrightarrow{R}\right)  =\frac{\exp\left(
-|\overrightarrow{R}-\overrightarrow{R}_{0}|\right)  }{\sqrt{|\overrightarrow
{R}-\overrightarrow{R}_{0}|^{2}+\varkappa^{2}}} \label{Finst}%
\end{equation}
where the parameter $\varkappa$ should be found from the minimization of the
action $S_{eff}$ (\ref{Seff1}) calculated on this configuration. We found that
$\varkappa=\left(  \frac{3}{4}\right)  ^{2/3}$. The effective action $S_{eff}$
for this value of $\varkappa$ is%

\[
S_{eff}=\left(  \frac{|\mu|}{E_{0}}\right)  ^{1/2}A
\]
where%

\begin{equation}
A=\frac{1}{2^{5/2}}\int d^{3}R\left(  L^{inst}\right)  ^{2}\sim4 \label{A1}%
\end{equation}
Finally we get the following estimation for the correlator%

\begin{equation}
<\overline{\Psi}\left(  \overrightarrow{r},t\right)  \Psi\left(
\overrightarrow{r},t\right)  >=n_{c}^{\left(  0\right)  }\int d^{3}%
R_{0}\left(  f^{inst}\left(  \overrightarrow{R}\right)  \right)  ^{2}%
\exp\left\{  -\left(  \frac{|\mu|}{E_{0}}\right)  ^{1/2}A\right\}  \label{nZM}%
\end{equation}

Hence the total number of particles%

\begin{align}
N  &  =n_{c}^{\left(  0\right)  }\int d^{3}r\int D\overrightarrow{R}%
_{0}\left(  f^{inst}\left(  \overrightarrow{R}\right)  \right)  ^{2}%
\exp\left\{  -\left(  \frac{|\mu|}{E_{0}}\right)  ^{1/2}A\right\}
=\label{N1}\\
&  =\Omega n_{c}^{\left(  0\right)  }\exp\left\{  -\left(  \frac{|\mu|}{E_{0}%
}\right)  ^{1/2}A\right\} \nonumber
\end{align}
Thus using Eqs. (\ref{N1}) and (\ref{nc}) we get the equation for the chemical
potential, which is%

\begin{equation}
\frac{ng}{|\mu|}=\exp\left\{  -\left(  \frac{|\mu|}{E_{0}}\right)
^{1/2}A\right\}  \label{EqCP}%
\end{equation}

Strictly speaking, this equation is valid for the large values of the chemical
potential so far as the saddle point calculation of the functional integral
(\ref{PsiPsi}) is valid for $|\mu|>>E_{0}$. However, we think that, as one can
guess from the form of the effective action (\ref{Seff1}), the equation of
this form is valid for $|\mu|\sim ng\sim E_{0}$ too. In this case the
effective action is dimensionless and does not depend on any parameter. The
main contribution to the functional integral (\ref{PsiPsi}) in this case
provided by the field $f$ with the asymptotic (\ref{e0R}), as as it is for
instantons in the saddle point approximation. This contribution should be of
the order of unity, so the estimation for the chemical potential is $|\mu|\sim
ng\sim E_{0}$. Eq. (\ref{EqCP}) has a solution for the densities smaller than
the critical density $n<n_{c}$, and does not have a solution for $n>n_{c}$,
where $n_{c}\sim\frac{E_{0}}{g}$. Formally, solution of Eq. (\ref{EqCP}), if
exists, has two branches (see Fig.1 for illustration). But the situation is
different from the first order phase transitions, since only one of them is
stable. The second solution has negative compressibility $\frac{\partial\mu
}{\partial n}<0$, and should be rejected. In a sense, the situation recalls a
bifurcation point rather a first order phase transition. The stable solution
of (\ref{EqCP}) with respect to $\mu$ as a function of $n$, in the region
$|\mu|>>E_{0}$ looks like%

\begin{equation}
\mu=-E_{0}\ln^{2}\left(  \frac{E_{0}}{ng}\right)  \label{mu0}%
\end{equation}
%

\begin{figure}
[ptb]
\begin{center}
\includegraphics[
height=3.1967in,
width=3.5034in
]%
{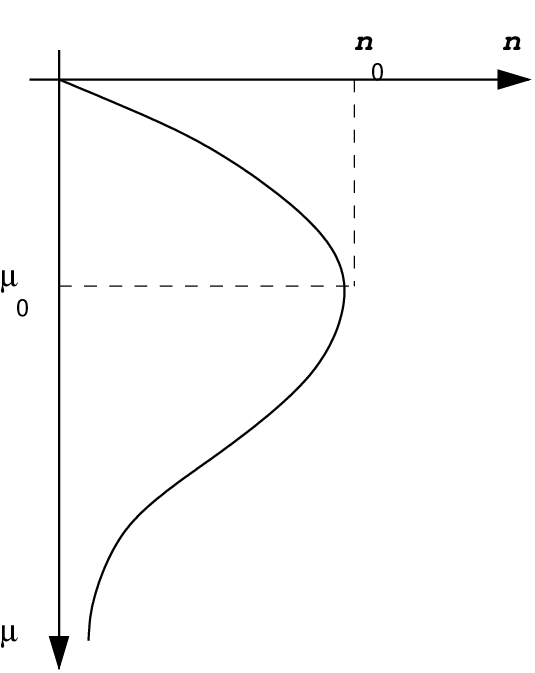}%
\end{center}
\end{figure}

Fig.1 Chemical potential as a function of density n.

It coincides with the expression for the chemical potential obtained recently
in \cite{Pokr}, \cite{Pokr1}. The value of the chemical potential $|\mu|$ for
densities $n\lesssim n_{c}$ can be estimated as $\mu\sim\mu_{c}$, where
$\mu_{c}\sim-E_{0}$. When the density $n$ becomes more than $n_{c}$ there is
no solution of Eq. (\ref{EqCP}), which means that for $n>n_{c}$ the chemical
potential will change the sign to positive and the system becomes delocalized
and superfluid.

Note that the exponential in Eq. (\ref{N1}) is typical for the tail of the
density of states in 3D for the negative energy of particles in a random potential.

In order to understand long distance correlation properties of the system in
the regime $n<n_{c}$ consider the correlator $<\overline{\Psi}\left(
\overrightarrow{r}_{1},t\right)  \Psi\left(  \overrightarrow{r}_{2},t\right)
> $ for $\mid\overrightarrow{r}_{1}-\overrightarrow{r}_{2}\mid\rightarrow
\infty$. The functional integral in the form of the new variables is%

\[
<\overline{\Psi}\left(  \overrightarrow{r}_{1}\right)  \Psi\left(
\overrightarrow{r}_{2}\right)  >=\int DfDe\exp\left\{  -S_{eff}\right\}
f\left(  \overrightarrow{r}_{1}\right)  f\left(  \overrightarrow{r}%
_{2}\right)  \overline{e}\left(  \overrightarrow{r}_{1}\right)  e\left(
\overrightarrow{r}_{2}\right)
\]

In the saddle point approximation the configurations of fields $\left(
f\left(  \overrightarrow{r}_{1}\right)  \text{, }e\left(  \overrightarrow
{r}_{1}\right)  \right)  $, $\left(  f\left(  \overrightarrow{r}_{2}\right)
\text{, }e\left(  \overrightarrow{r}_{2}\right)  \right)  $, when
$\mid\overrightarrow{r}_{1}-\overrightarrow{r}_{2}\mid\rightarrow\infty$,\ are
described by the instantons localized near far distant points $\overrightarrow
{r}_{1}$ and $\overrightarrow{r}_{2}$, respectively, and the correlator
$<\overline{\Psi}\left(  \overrightarrow{r}_{1}\right)  \Psi\left(
\overrightarrow{r}_{2}\right)  >$ is exponetially small, i.e. the system has
no long range order for densities $n<n_{c}$. Note that the correlator
$<\overline{e}\left(  \overrightarrow{r}_{1}\right)  e\left(  \overrightarrow
{r}_{2}\right)  >$ is equal to zero for $\mid\overrightarrow{r}_{1}%
-\overrightarrow{r}_{2}\mid\rightarrow\infty$ due to the independence of the
phases of the different instantons.

\section{Discussion}

The effective action for many-particle system in random potential is derived
in the assumption of weak interaction between particles but without
restrictions on the strength of the random field fluctuations. The approach we
develop is different from the usual Q-Lagrangian approach used in the theory
of weak localization. The order parameter which could characterize the
transition from localized to delocalized state has another structure and
characterized by the "classical" field $\Psi$. We obtain the effective action
by the integration over the random external field at the first stage,
following by integration over the quantum fluctuations described by the field
$\psi$. The approximation we use for the integration over the field $\psi$ is
correct in quasi classical case as well as for small density of the system or
weak interaction between particles. In this paper we consider the properties
of Bose gas in random external potential with the Gaussian correlator for the
random field fluctuations in the case of the negative chemical potential,
i.e., in the region of the strong localization. The obtained effective action
has instantons which are localized in space and give the finite value for the
effective action. The contribution of these instantons to the functional
integral (\ref{PsiPsi}) defines the equation for the chemical potential as a
function of the averaged density. The solution of this equation exists only if
the averaged density is smaller than the critical value $n_{c}\sim\frac{E_{0}%
}{g}$. The chemical potential for the densities $n\sim n_{c}$\ is of the order
$\mu\sim\mu_{c}=-E_{0}$.

We have shown that the system has no long range order for the densities
$n<n_{c}$, whilst the chemical potential becoms positive, and the strong
localized state transfers to the delocalized superfluid state.

One of the main important problems remaining for future is a more careful
semiclassical investigation of instantonic solutions found here for different
space dimensions. It would be also interesting to apply the developed method
to the regime $\mu\sim\mu_{c}$, where it could be possible to see the phase
transition, probably of the first order, when $\mu$ changes the sign.

Another problem where the described path integral Keldysh-Schwinger technique
can be applied, is fermionic particles in a random potential. Here higher
order powers of electron-electron interaction cannot be neglected, and the
full action should be treated non perturbatively. It seams reasonable to
concentrate on the 1+1 dimensional space case for this problem and try to
analyze a relation of the found chiral Lagrangian to solvable sigma model cases.

\section{Acknowledgments}

A.Babichenko is thankful to Einstein Center of Weizmann Institute. His work
was also partially supported by ISF grant 286/04.


\begin{thebibliography}{99}                                                                                               %


\bibitem {Berez}V.L. Berezinskii, JETP, 65, 125 (1973).

\bibitem {And1}E. Abrahams, P.W. Anderson, D.C. Licciardello, T.V.
Ramakrishnan, Phys. Rev. Lett. 42, 673 (1978).

\bibitem {Lee}A. Lee, T.V. Ramakrishnan, Rev. Mod. Phys. 57, 287 (1985).

\bibitem {And2}P.W. Anderson, Phys. Rev. 109, 1492 (1958).

\bibitem {Lif}I.M. Lifshitz, Sov. Phys. JETP 26, 462, (1968).

\bibitem {Zit}J. Zittartz, J. Langer, Phys. Rev. 148, 741 (1966).

\bibitem {Halp}B.I. Halperin, and M. Lax, Phys. Rev. 153, 802 (1966).

\bibitem {Lif1}I.M. Lifshitz, S.A. Gredeskul, and L.A. Pastur, "Introduction
to the theory of the disordered systems", Wiley Interscience, New York 1988.

\bibitem {Weg}F. Wegner, Z. Phys. B 35, 207 (1979).

\bibitem {Gor}L.P. Gorkov, A.I. Larkin, D.E. Khmelnitskii, Sov. Phys. JETP
Lett. 30, 228 (1979).

\bibitem {Ef}K.B. Efetov, Adv. Phys. 32, 53 (1983).

\bibitem {Alt}B.L. Altshuller, A.G. Aronov, Sov. Phys. JETP 50, 968 (1979).

\bibitem {Fink}A.M. Finkelstein, Sov. Phys. JETP, 57, 97 (1983).

\bibitem {Ef1}I.L. Aleiner, K.B. Efetov, Phys. Rev. B 76, 075102 (2006);
cond/mat 0602309.

\bibitem {Schw}J. Schwinger, J. Math. Phys. 2, 407 (1961).

\bibitem {Keld}L. V. Keldysh, JETP 47, 1515 (1964).

\bibitem {BK}V.S. Babichenko, and A.N. Kozlov, Solid State Comm. 59, 39 (1986).

\bibitem {Kam}A. Kamenev and A. Andreev Phys. Rev. B 60, 2218 (1999); A.
Kamenev, cond-mat/0412296.

\bibitem {LLP}L.D. Landau, E.M. Lifshits, L. P. Pitaevskii, "Physical
Kinetics", Nauka 1979.

\bibitem {Bel}S.T. Belyaev, JETP 34, 417 (1958).

\bibitem {Fisher}M.P.A. Fisher, P.B. Weichman, G. Grinstein, D.S. Fisher,
Phys. Rev. B 40, 546 (1989).

\bibitem {Huang}K. Huang, H.F. Meng, Phys. Rev. Lett. 69, 644 (1992).

\bibitem {Pit}S. Giorgini, L. Pitaevskii, and S. Stringari, Phys. Rev. B 49,
12938 (1994); cond-mat/9402015.

\bibitem {Vin}A.V. Lopatin, V.M. Vinokur, Phys. Rev. Lett. 88, 235503 (2002).

\bibitem {Grah}G.M. Falco, A. Pelster, R. Graham, Phys. Rev. A 75, 063619 (2007).

\bibitem {Pokr}G.M. Falco, T. Nattermann, and V.L. Pokrovsky, Phys. Rev. Lett.
100, 060402 (2008).

\bibitem {Pokr1}G.M. Falco, T. Nattermann, and V.L. Pokrovsky, cond-mat/0808.2565v1.

\bibitem {Reppy}J.D. Reppy, J. Low Temp. Phys. 87, 205 (1992).

\bibitem {Vicente}C.L. Vicente, et. al, Phys. Rev. B 72, 094519 (2005).

\bibitem {Lye}J.E. Lye, et.al., Phys. Rev. Lett. 95, 070401 (2005).

\bibitem {Schulte}T. Schulte, et.al, Phys. Rev. Lett. 95, 170411 (2005).

\bibitem {Fallani}L. Fallani, et.al, Phys. Rev. Lett. 98, 130404 (2007).

\bibitem {Lugan}P. Lugan, et.al, Phys. Rev. Lett. 98, 170403 (2007).

\bibitem {San}L. Sanchez-Pelencia, et.al, Phys. Rev. Lett. 98, 210401 (2007).

\bibitem {Chen}Y.P. Chen, et.al, Phys.Rev. A 77, 033632 (2008).

\bibitem {Billy}J. Billy, et.al, Nature 453, 891 (2008).
\end{thebibliography}
\end{document}